\documentclass[aps,prl,reprint,superscriptaddress,longbibliography]{revtex4-2}

\usepackage{amsmath}
\usepackage{amssymb}
\usepackage{graphicx}
\usepackage{bm} 
\usepackage[colorlinks=true,citecolor=blue,linkcolor=blue,urlcolor=blue]{hyperref}

\begin{document}

\title{Intuition emerges in Maximum Caliber models at criticality}

\author{Lluís Arola-Fernández}
\email{lluis.arolaf@urv.cat}

\affiliation{Instituto de Física Interdisciplinar y Sistemas Complejos IFISC (CSIC-UIB), 
Campus UIB, 07122 Palma de Mallorca, Spain}

\affiliation{Departament d'Enginyeria Informàtica i Matemàtiques, Universitat Rovira i Virgili, 
43007 Tarragona, Catalonia, Spain}

\altaffiliation{Current address.}

\date{\today}

\begin{abstract}
Whether large predictive models merely parrot their training data or produce genuine insight lacks a physical explanation. This work reports a primitive form of \emph{intuition} that emerges as a metastable phase of next-token prediction under future path-entropy maximization. The intuition mechanism is discovered via \emph{mind-tuning}, the minimal principle that imposes Maximum Caliber in predictive models with a control temperature-like parameter~$\lambda$. Training on random walks in deterministic mazes reveals a rich phase diagram: \emph{imitation} (low~$\lambda$), rule-breaking \emph{hallucination} (high~$\lambda$), and a fragile in-between window exhibiting strong protocol-dependence (hysteresis) and multistability, where models spontaneously discover novel goal-directed strategies. These results are captured by a mechanistic low-dimensional theory and frame intuition as an emergent property at the critical balance between memorizing what is and wondering what could be.
\end{abstract}

\maketitle

\begin{figure}[h!]
\centering
\includegraphics[width=0.32\columnwidth]{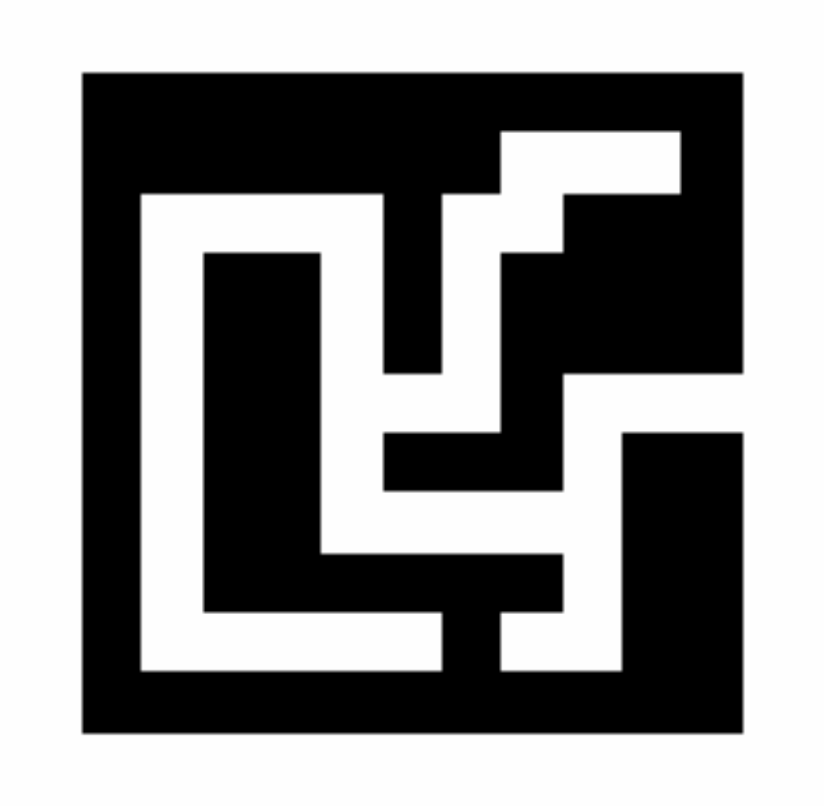}
\caption{\textbf{\emph{Gedankenexperiment} on emergent reasoning.} A minimal environment abstracts a reasoning task into its essential components: a constrained space (a maze) and a hidden, optimal solution (to escape). The reader's own intuition immediately grasps the task, yet a standard predictive model trained on random walk trajectories (i.e., non-intelligent data without rewards) will never discover it.}
\label{fig:conceptual}
\end{figure}

\textit{Introduction.---} The rise of large-scale predictive models is reshaping artificial intelligence and transforming science and society. This progress is built upon a dominant scaling paradigm: pre-training autoregressive neural networks~\cite{Vaswani2017} with enormous parameter counts on big volumes of data~\cite{Kaplan2020} using massive compute resources~\cite{Hoffmann2022}. When coupled with powerful search at inference time~\cite{DeepSeek2025}, this approach has yielded impressive performance in complex games~\cite{Shojaee2024}, medical diagnosis~\cite{Brodeur2024} and algorithmic discovery~\cite{AlphaEvolve2025}. Yet, the brute-force solution does not match the elegant efficiency of natural intelligence, which discovers intuitive shortcuts and novel, creative strategies from sparse data without rewards~\cite{Chollet2019}. This contrast sharpens a foundational debate: are these models showing sparks of artificial general intelligence (AGI)~\cite{Bubeck2023}, or are they ``stochastic parrots''~\cite{Bender2021} that leverage vast experience to create an illusion of thought~\cite{Shojaee2024, Mitchell2023}? While often addressed via complex reasoning benchmarks~\cite{Liang2022}, the paradigm's limits can be distilled into a simple \emph{Gedankenexperiment} (Fig.~\ref{fig:conceptual}).

This work provides a physical explanation for this leap. We introduce \emph{mind-tuning}, a simple principle that balances next-token prediction against future path-entropy maximization with a temperature-like parameter~$\lambda$. To our knowledge, mind-tuning is the minimal implementation of the Maximum Caliber (MaxCal) principle \cite{Jaynes1980, Presse2013,dixit2018} compatible with autoregressive training. It reveals the emergence of a fragile metastable phase, within a narrow temperature window between \emph{imitation} and \emph{hallucination} regimes, that is reminiscent of \emph{intuition}.

While our intuition mechanism points toward a horizon of diverse futures to explore, the prevailing paradigm remains blind, fixated only on predicting the next token. Constrained path-entropy maximization is already implicit in \emph{intrinsic motivation} frameworks~\cite{Kiefer2025} like Causal Entropic Forces~\cite{WissnerGross2013}, Active Inference~\cite{Wen2025}, Empowerment~\cite{Klyubin2005}, or the Maximum Occupancy Principle~\cite{Ramirez2024}. Yet, a physical basis for such emergent behavior in pure predictive models has remained elusive. The metastable regime reported here, bounded by distinct, entropy and energy-driven transitions with strong hysteresis and multistability, explains that emergent reasoning is both rare and protocol-dependent. Furthermore, the high-dimensional mechanisms behind this phenomenology are captured analytically by a low-dimensional theory. 

This perspective casts intelligence as a state of computational matter \cite{Friston2022}, building on a rich history of minimal models for emergent cognitive behavior, from Hopfield's memory \cite{Hopfield1982} and Kuramoto's synchronization~\cite{Kuramoto1975} to phenomena in deep learning like double-descent~\cite{Belkin2019}, grokking~\cite{Power2022}, neural collapse~\cite{Papyan2020}, symmetry-breaking~\cite{Ziyin2025symmetry}, and collective learning~\cite{Arola2024}, often analyzed through spin-glass analogies~\cite{Carleo2019} and phase diagrams~\cite{lewkowycz2020,Arola2024}. The phase-transition picture is key to research showing that intelligent systems may operate near a critical point, at the ``edge of chaos"~\cite{Munoz2018,zhang2025,jimenez2025}. At criticality, fluctuations and system responsiveness peak~\cite{Munoz2018,Arola2020}, creating the ideal conditions for the leap from mimicry to insight. In the learning problem, our theory points toward a critical scaling axis driven by the system's intrinsic dynamics and suggests that current models operate in a suboptimal imitation phase, lacking the intuition that a physical mechanism unlocks.

\textit{Mind-tuning.---} We focus on reasoning problems solvable by generating trajectories $z=(x_0,a_0,x_1,a_1,\dots)$. The system's behavior is governed by a policy $\pi_{\theta,\beta}$, a neural network with parameters $\theta$ that maps a data history $h_t=(x_0,a_0,\dots,x_{t})$ to a probability distribution over a discrete set of actions $\mathcal{A}$ via a softmax function
\begin{equation}
\label{eq:policy_softmax}
\pi_{\theta,\beta}(a_t\!\mid\! h_t)=\frac{e^{\beta\,\ell_\theta(h_t,a_t)}}{\sum_{a'\in\mathcal{A}}e^{\beta\,\ell_\theta(h_t,a')}},
\end{equation}
where $\ell_\theta$ are the network's output logits and $\beta$ controls the policy's stochasticity. This general setting includes state-decision spaces, standard autoregressive models where histories contain tokens and other representations (see SM Sec. S1 for implementation details).

To isolate the intuition mechanism, we assume an offline, imperfect setting \cite{levine2020}: the model never interacts with the environment, has no external rewards, and learns from a dataset $\mathcal{D}$ of non-optimal histories. How can a purely predictive model discover a better solution than what it has seen? By biasing prediction toward futures with high causal path diversity, as prescribed by the Maximum Caliber (MaxCal) principle \cite{Jaynes1980}: \emph{among all dynamics consistent with known constraints, prefer those that maximize the entropy of trajectories}.

The most unbiased learning objective that imposes MaxCal is the free-energy-like functional
\begin{equation}
\label{eq:F_general}
\mathcal{F}_{\lambda,\beta,\tau}(\theta) = \mathcal{E}_\beta(\theta) - \lambda \mathcal{H}_{\tau,\beta}(\theta),
\end{equation}
where $\lambda\!\ge\!0$ is an effective temperature controlling the energy–entropy trade-off. The first term is the standard Cross-Entropy or negative log-likelihood $(\mathcal{E}$), measuring the cost of imitating the training data
\begin{equation}\label{eq:CE_def}
\mathcal{E}_\beta(\theta) = \left\langle -\log\pi_{\theta,\beta}(a_t|h_t) \right\rangle_{(h_t, a_t) \in \mathcal{D}}.
\end{equation}
This energy $\mathcal{E}$ is traded against the causal path-entropy $\mathcal{H}$, a Shannon entropy of self-generated futures up to a horizon of length $\tau$
\begin{equation}\label{eq:H_general}
\mathcal{H}_{\tau,\beta}(\theta) = \left\langle \frac{1}{\tau}\left\langle-\ln P(z_{\text{future}}|h_{t})\right\rangle_{z_{\text{future}} \sim \pi_{\theta,\beta}} \right\rangle_{h_{t} \in \mathcal{D}}.
\end{equation}
Eq.(\ref{eq:H_general}) is estimated over the cone of futures induced by the model itself (see SM Sec. S2B for entropy calculations), making the objective function inherently subjective and self-referential, as the internal beliefs dynamically shape the learning landscape. The gradient update
\begin{equation} \label{eq:forces}
\theta(t+1) \leftarrow \theta(t) + \eta [{-\nabla_\theta \mathcal{E}_\beta(\theta)} + {\lambda \nabla_\theta \mathcal{H}_{\tau,\beta}(\theta)}]
\end{equation}
frames learning as a competition between prediction and causal entropic forces acting on the system's degrees of freedom, i.e. the network weights. To our knowledge, this self-contained mechanism is the minimal MaxCal implementation compatible with prevalent offline auto-regressive training. Unlike surprise-minimization~\cite{Heins2024,Friston2010}, here the entropic term rewards keeping plausible futures open, pulling toward the \emph{adjacent possible}~\cite{Kauffman2000}, without environment interaction~\cite{Klyubin2005,Ramirez2024,eysenbach2022}. The framework also admits a Bayesian interpretation \cite{Jaynes1957,Zdeborova2016}: standard auto-regressive training use flat priors. In mind-tuning, instead, the data likelihood filters an optimistic entropic prior over futures with high diversity.

\begin{figure}[b!]
    \centering
    \includegraphics[width=\columnwidth]{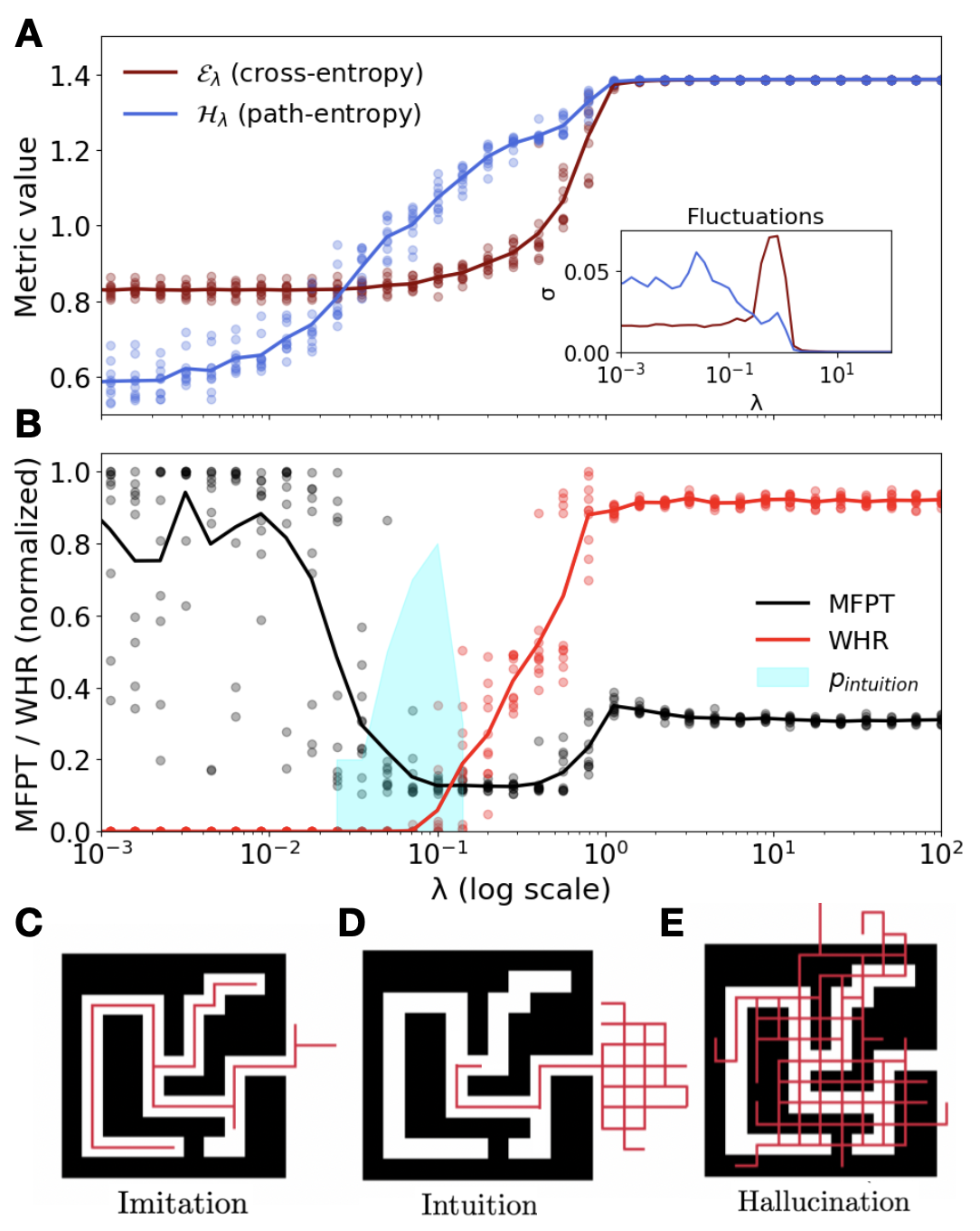}
    \caption{\textbf{Experimental phase diagram.} Sweeping $\lambda$ reveals three behavioral phases. (A) Genotype metrics: Cross-Entropy ($\mathcal{E}$) and causal path-entropy ($\mathcal{H}$). Inset: steady-state fluctuations $\sigma$ over different initial realizations depending on $\lambda$. (B) Phenotype metrics: Mean First Passage Time (MFPT), Wall Hit Ratio (WHR) and intuition likelihood (see SM Sec. 4B). (C-E) Example trajectories for each phase: (C) Imitation, (D) Intuition, and (E) Hallucination.}
    \label{fig:diagrams}
\end{figure}

\textit{Experiments.---} We test this principle in the minimal sandbox of the \emph{Gedankenexperiment} (Fig.~\ref{fig:conceptual}). A model is trained on constrained random-walk trajectories, which respect the maze walls but contain no intelligent strategies for escaping. Sweeping the parameter $\lambda$ yields a rich phase diagram, with clear transitions in both genotype (Fig.~\ref{fig:diagrams}A) and phenotype (Fig.~\ref{fig:diagrams}B) metrics.

For low $\lambda$, the system is in an \emph{imitation} phase: cross-entropy is low, path-entropy is low, and trajectories reproduce the suboptimal random walks from the data, leading to a high Mean First Passage Time (MFPT) to the exit (Fig.~\ref{fig:diagrams}C). For high $\lambda$, the entropic term dominates and the system enters a \emph{hallucination} phase: cross- and path-entropy are high; maze rules are broken to maximize path diversity, and the Wall Hit Ratio (WHR) increases sharply (Fig.~\ref{fig:diagrams}E). Between these two regimes lies a narrow \emph{intuition} phase, where the trade-off between $\mathcal{E}$ and $\mathcal{H}$ yields an emergent strategy: the model discovers the shortest legal path to the exit (Fig.~\ref{fig:diagrams}D), achieving minimal MFPT with zero WHR. The separation between the fluctuation peaks of $\mathcal{E}$ and $\mathcal{H}$ (Fig.~\ref{fig:diagrams}A inset) reveals distinct entropy- and energy-driven phase boundaries.

\begin{figure}[b!]
    \centering
    \includegraphics[width=\columnwidth]{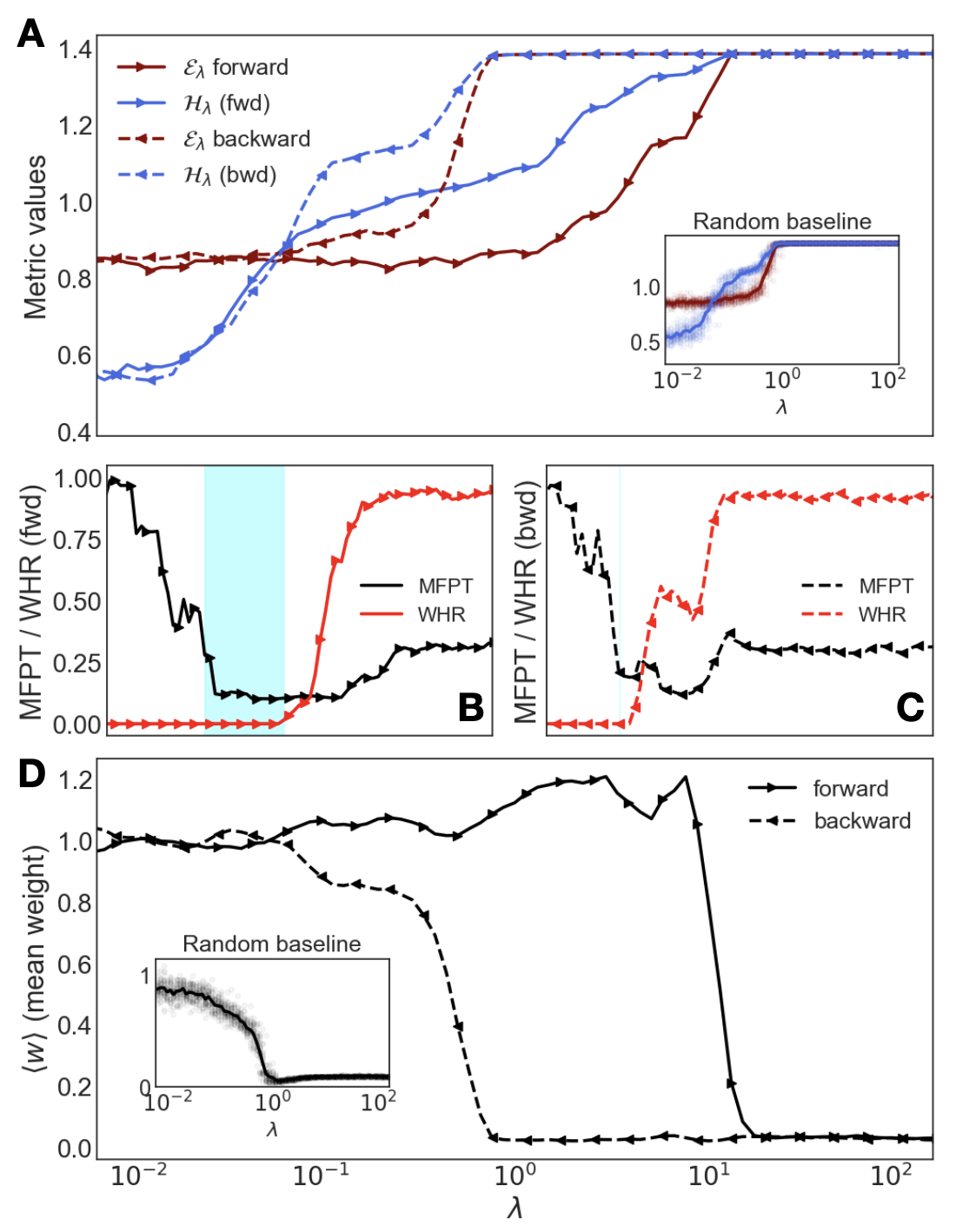}
    \caption{\textbf{Hysteresis and protocol-dependence.} Comparing a forward (solid) and backward (dashed) sweep of $\lambda$ reveals that the intuitive state is stable once found. (A) Hysteresis loop in genotype metrics ($\mathcal{E}, \mathcal{H}$). (B, C) Phenotype for the forward and backward sweeps, respectively, with the forward sweep showing a wider intuition window. (D) The mean network weight $\langle w \rangle$ acts as an order parameter capturing the system's bistability. Insets show baselines without protocol.}
    \label{fig:protocols}
\end{figure}

Operationally, this critical learning phase maximizes future path-entropy with minimal cross-entropy, enabling novel, goal-directed behavior at inference without interaction or explicit rewards. Reaching this phase depends on data quality and model complexity, requiring a sufficiently large future horizon and adequate model capacity (see SM Sec. S3 for a parametric study). The fragility of the mechanism is tied to multistability, as observed when applying adiabatic protocols that smoothly sweep the control parameter $\lambda$ (Fig.~\ref{fig:protocols}). A large hysteretic loop appears in the genotype metrics (A), which has behavioral consequences in the phenotype: while a forward sweep from $\lambda \approx 0$ opens the intuition window, with low MFPT and low WHR (B), a backward sweep starting from high $\lambda$ does not reach the desired phase (C). The bistability is captured by an effective order parameter --the mean network weight-- which remains in an ordered intuitive state once the system has been guided there (D). The adiabatic protocol shows that a self-referential fine-tuning from imitation to controlled imagination allows the system to stabilize in a metastable phase, a process that motivates the term \emph{mind-tuning}.

\textit{Effective theory.---} The phenomenology of mind-tuning emerges from a high-dimensional, multistable free-energy landscape. We capture the essential mechanism in a scalar order parameter $m\in[0,1]$, representing the model’s rationality, and define a Boltzmann policy with an effective potential $U_m(a)$:
\begin{equation}
    p_{m,\beta}(a|h_t) = \frac{e^{-\beta U_m(a)}}{\sum_{a'\in\mathcal{A}} e^{-\beta U_m(a')}}.
\end{equation}
Actions, or decisions, are classified into optimal $a^*$, rational-but-suboptimal $a^r$, and non-rational $a^n$ and $m_D$ is a free parameter representing the training data's rationality. The effective costs,
\begin{align}
    U(a^*) &= 0, \nonumber\\
    U(a^r) &= \frac{\max(0, m - m_D)}{1 - m_D}, \\
    U(a^n) &= m, \nonumber
\end{align}
are designed to create a trade-off: as the model's rationality $m$ improves beyond the data's, the cost of suboptimal-but-legal actions grows, forcing a choice between true optimality and rule-breaking. For the simple Markovian maze with a small state-space, the free energy $\mathcal{F}_\lambda(m)$ can be computed analytically (see SM Sec. S4A). For a given $\lambda$, one can also explore the learning dynamics in this landscape by sampling rationality states $m$ from the equilibrium distribution $P(m) \propto e^{-\hat{\beta}\mathcal{F}(m)}$, where the inverse temperature $\hat{\beta}$ controls the exploration-exploitation trade-off, modeling stochasticity during gradient descent.  

\begin{figure}[t!]
    \centering
    \includegraphics[width=\columnwidth]{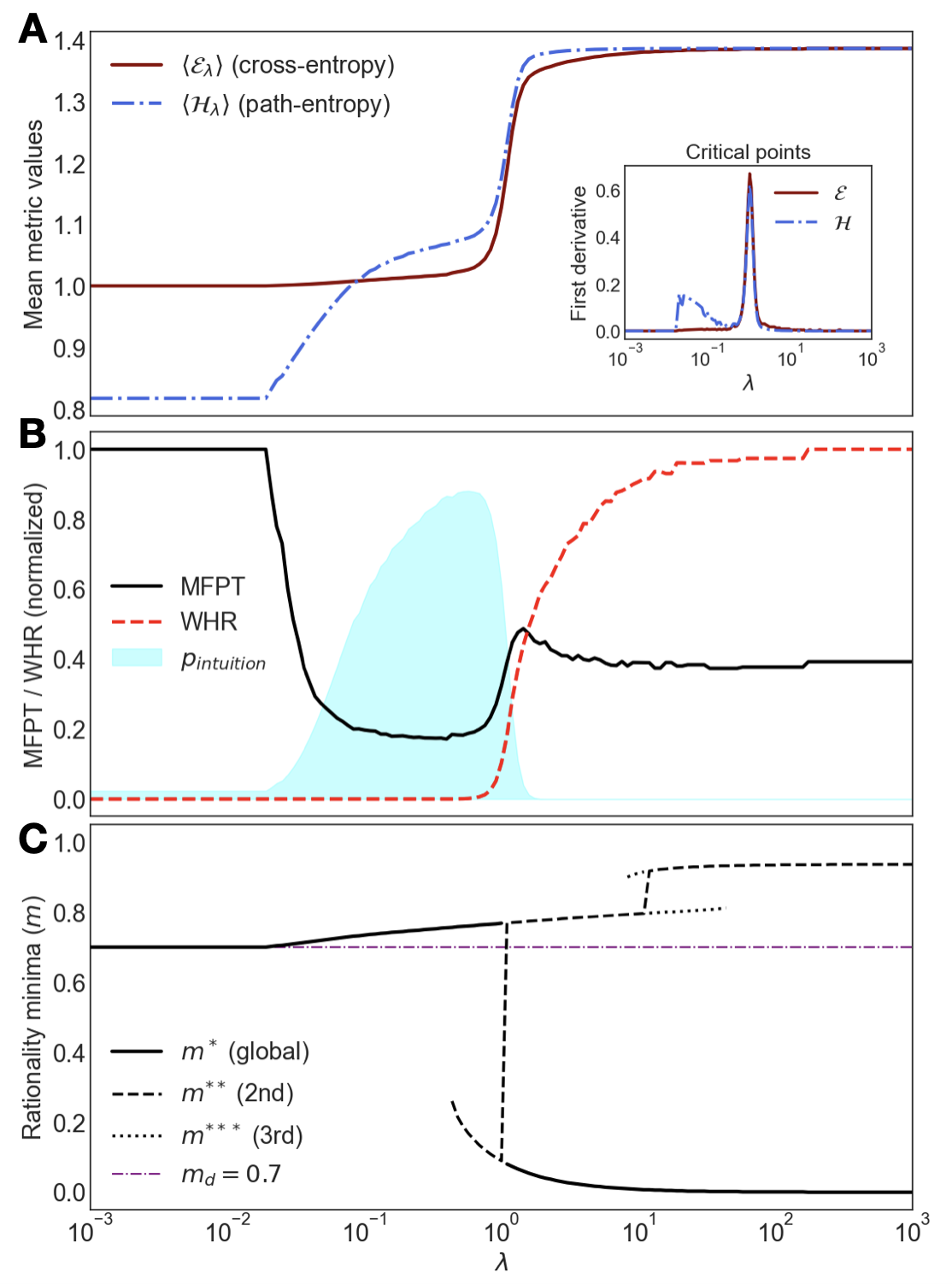}
    \caption{\textbf{Theoretical predictions.} The low-dimensional model reproduces the experimental findings. (A) Theoretical $\mathcal{E}$ and $\mathcal{H}$ vs. $\lambda$. (B) Corresponding MFPT and WHR. (C) Minima of the free-energy landscape vs the control parameter $\lambda$. The plot reveals coexisting stable states ($m^*, m^{**},m^{***}$) and a first-order transition where the global minimum jumps discontinuously, explaining the observed hysteresis.}
    \label{fig:mechanism}
\end{figure}

This effective theory qualitatively reproduces the experimental phase diagram, including the transitions in both genotypic (Fig.~\ref{fig:mechanism}A) and phenotypic metrics (Fig.~\ref{fig:mechanism}B). The underlying mechanism is revealed by exploring the minima of the free-energy landscape, found by solving $\partial\mathcal{F}_\lambda(m)/\partial m = 0$. This analysis confirms a smooth, entropy-driven transition followed by an abrupt, first-order energy-driven one, creating a bistable region where intuition ($m>m_D$) and hallucination ($m\ll m_D$) coexist (Fig.~\ref{fig:mechanism}C). Intriguingly, the theory further predicts a more elusive \emph{inspiration} phase: a third stable solution with $m\approx 1$, associated to a state of true creative insight. This strategy abruptly departs from data and represents internalized understanding. Unlike the subtle intuitive state, which often requires a high inference $\beta$ to be executed without error, this inspired solution would be robust even with a noisy policy. Yet, it is hidden within a tiny basin of attraction masked by the dominant hallucination phase (see SM Sec. S4.C). These predictions point to a very rich phase diagram, where intuition may be the trigger of even more exotic phenomena.

Accessing these different cognitive phases requires navigating a complex landscape. Indeed, the observed hysteresis and the success of the adiabatic protocol are explained by this multi-stability. The analytical phase-diagram (Fig.~\ref{fig:mechanism}C) shows that slowly increasing $\lambda$ is a safe route to guide the system into the \emph{intuition} basin of attraction. In Bayesian terms, it first grounds the model with the data likelihood before introducing the entropic prior. Reaching more exotic phases in the landscape, like the predicted inspiration state, would likely demand more complex, non-equilibrium protocols.

\textit{Discussion.---} High-quality human data can carry an implicit drive toward path diversity, and optimization itself can induce entropic pressures that improve generalization~\cite{Ziyin2025Entropic}, yielding an ``intelligent simulator'' from curated experience. This view predicts that current models should spontaneously increase their causal path entropy with scale. Our framework makes this drive explicit and grounded in MaxCal, providing a shortcut to intuition that encodes implicit search into model weights to reduce the need for expensive search at inference \cite{Belcak2025}. These results point toward a hidden axis, \emph{training-time imagination}, that may be key to unlock out-of-distribution generalization in offline predictive models \cite{levine2020}.

Our results are demonstrated in a minimal sandbox, a choice that is deliberate. The maze is the simplest non-trivial setting where the mechanism can be isolated and reproduced analytically. Many reasoning tasks can be viewed as navigation through a ``conceptual maze'' where a key insight unlocks a vastly larger state-space \cite{WissnerGross2013,Klyubin2005,Ramirez2024,Friston2022}. This argument promises applications in control \cite{WissnerGross2013,Ramirez2024}, reasoning~\cite{Chollet2019,Wang2024}, and planning \cite{Wang2024}. Stefan Zweig’s \emph{The Royal Game}~\cite{Zweig1943} provides a compelling literary analogue: a prisoner achieves chess mastery by first studying games (imitation) and then playing against himself in his mind (imagination). His triumph occurs at the edge of madness, a state mirroring intuition coexisting with hallucination in our phase diagram.

Yet, scaling mind-tuning to real-world cases faces significant challenges. Computationally, estimating path-entropy for long horizons is hard due to the combinatorial explosion of futures \cite{Jaynes1980}. This requires designing clever sampling strategies \cite{WissnerGross2013,Aguilar2022}, perhaps inspired by dreaming, hierarchical reasoning \cite{Wang2024} and unconventional methods and architectures \cite{Labay2025,brunner2025}. Theoretically, a full characterization of the phase diagrams and universality classe is needed to design optimal tuning protocols~\cite{manzano2024}. For uncharted domains, identifying the right spaces for entropy maximization can be difficult and the offline theory may need data augmentation from environment interaction~\cite{Wen2025}. Yet, tuning $\lambda$ for future diversity in practice can turn into an alignment problem, trading benefits for safety \cite{Arenas2011}. Despite these challenges, this work takes a high-risk, high-reward route to reframing intelligence not merely as compression and computation, but as a physical phenomenon emerging at criticality.

\textit{Acknowledgments.---} The author thanks many colleagues at IFISC and URV for enriching discussions. This work has been partially
supported by the María de Maeztu project CEX2021-001164-M funded by the MICIU/AEI/10.13039/501100011033 and by Programa Maria Goyri URV.

\clearpage
\onecolumngrid
\appendix
\begin{center}
\textbf{\large Supplementary Material for: ``Intuition emerges in Maximum Caliber models at criticality"}
\end{center}
\setcounter{equation}{0}
\setcounter{figure}{0}
\setcounter{table}{0}
\setcounter{page}{1}
\makeatletter
\renewcommand{\theequation}{S\arabic{equation}}
\renewcommand{\thefigure}{S\arabic{figure}}
\renewcommand{\bibnumfmt}[1]{[S#1]}
\renewcommand{\citenumfont}[1]{S#1}

\section{S1. Experimental Setup and Hyperparameters}
\label{sec:si_implementation}

The experimental setting is a minimal yet non-trivial environment for testing emergent reasoning. It consists of a deterministic $24 \times 24$ maze with periodic boundary conditions, where an agent must find the path to a designated exit. This controlled testbed provides a tractable state space for analyzing the learning dynamics. The agent's behavior is determined by a policy network that maps the current state (2D position $x_t$) to a probability distribution over the four cardinal actions: $\mathcal{A} = \{\text{Up, Down, Right, Left}\}$. For auto-regressive training, a simple deterministic function $f(x_t,a)$ maps the last action to the next state. 

The training dataset $\mathcal{D}$ is intentionally non-optimal. In our main experiments, it contains $N=100$ trajectories, each of length $T=60$ steps, generated by a constrained random walks. These walkers respect the maze walls (i.e., never collide with them) but otherwise move randomly, exhibiting no goal-directed behavior. This design ensures that the optimal exit strategy is not present in the training data, forcing the model to discover it.

The model parameters $\theta$ are optimized by minimizing the free-energy functional $\mathcal{F}_{\lambda,\beta,\tau}(\theta)$ (Eq. (2) in the main text) via the Adam optimizer. The results presented in the main text (Fig. 2) are averaged over 20 independent training runs, each with a different random weight initialization, to ensure statistical robustness. The key hyperparameters used in the main experiments are: a policy network structured as a multi-layer perceptron (MLP) with one hidden layer of 128 neurons and ReLU activation; a learning rate of $1 \times 10^{-3}$; 300 training epochs per $\lambda$ value; and a future horizon of $\tau=40$ steps in the entropy calculation.

The policy stochasticities are set to $\beta = 1$ for training, $\beta = 5$ for entropy calculation (imagination), and $\beta = 10$ at inference time. A high imagination $\beta$ (compared to the training $\beta$) is beneficial for discovering hidden solutions that maximize causal entropy (i.e., finding the exit) with a finite $\tau$ and sparse data. A high inference $\beta$ is necessary to induce intuitive behavior in practice. In the intuition phase, the agent finds a superior solution but must execute its policy quite deterministically to follow the optimal path in the minimum time.

For problems that are not Markovian or where the data representation does not contain full state information (e.g., data are sequences of moves or the agent only sees its local environment), a more advanced neural network is required. Transformers are the standard for modeling long, non-Markovian sequences of tokens. Our framework naturally extends to these sequential autoregressive architectures, albeit at the cost of more parameters and computational effort.

\textit{Code availability.---} PyTorch source code to reproduce the results of this paper is publicly available on GitHub: \href{https://github.com/mystic-blue/mind-tuning}{https://github.com/mystic-blue/mind-tuning}.

\section{S2. Calculation of Objective Functionals}
\label{sec:si_derivation}
The mind-tuning objective function $\mathcal{F}_{\lambda,\beta,\tau}(\theta) = \mathcal{E}_\beta(\theta) - \lambda \mathcal{H}_{\tau,\beta}(\theta)$ consists of two key terms. Below we detail their calculation.

\subsection{A. Cross-Entropy Estimation}
The cross-entropy term $\mathcal{E}_\beta(\theta)$, defined in Eq. (3) of the main text, measures the model's ability to imitate the training data. It is estimated by averaging the negative log-likelihood of the actions taken in the dataset $\mathcal{D}$ given the preceding histories:
\begin{equation}
\label{eq:E_estimation_sup}
\hat{\mathcal{E}}_{\beta}(\theta) = \frac{1}{|\mathcal{D}|} \sum_{(h_t, a_t) \in \mathcal{D}} [ -\log \pi_{\theta,\beta}(a_t|h_t) ]
\end{equation}
where $|\mathcal{D}|$ is the total number of state-action pairs in the training set. This term encourages the policy to assign high probability to the trajectories observed during training.

\subsection{B. Causal Path-Entropy: Analytic Calculation for Markovian Systems}
For systems with fully-observed, discrete, and reasonably small state spaces $\mathcal{V}$, such as our maze environment, the path-entropy can be computed analytically. Since the system is Markovian ($h_t = x_t$), we can define a policy-dependent transition matrix $M_{\pi}$. The element $(M_{\pi})_{x',x}$ gives the probability of transitioning from state $x$ to state $x'$ under the current policy $\pi_{\theta,\beta}$. Specifically, $(M_{\pi})_{x',x} = \sum_{a \in \mathcal{A}} \pi_{\theta,\beta}(a|x)\delta_{x',f(x,a)}$, where $f(x,a)$ is the deterministic function that returns the next state.

Given a starting state $x_{start}$, we can compute the probability distribution over future states $\vec{\rho}_k$ at any time step $k$ by evolving an initial occupancy vector (a point mass at $x_{start}$) via the recursion $\vec{\rho}_{k+1} = M_{\pi} \vec{\rho}_k$. The conditional path-entropy for a trajectory starting at $x_{start}$ is then the time-averaged Shannon entropy of the policy, weighted by the occupancy probability at each future state:
\begin{equation}
\label{eq:H_analytic_sup}
\mathcal{H}_{\tau,\beta}(\theta|x_{start}) = \frac{1}{\tau}\sum_{k=0}^{\tau-1}\sum_{x\in\mathcal{V}} (\rho_k)_x \left[-\sum_{a\in\mathcal{A}}\pi_{\theta,\beta}(a|x)\log\pi_{\theta,\beta}(a|x)\right].
\end{equation}
The total functional $\mathcal{H}_{\tau,\beta}(\theta)$ is the expectation of Eq.~\eqref{eq:H_analytic_sup} over all starting states in the training dataset $\mathcal{D}$. This entire calculation is fully differentiable with respect to the network parameters $\theta$, allowing for efficient gradient-based optimization. This exact method was used to produce all experimental and theoretical results in this work. Its primary computational cost scales with the size of the state space $|\mathcal{V}|$, making it suitable for our testbed.

\subsection{C. Causal Path-Entropy: Monte Carlo Estimation for High-Dimensional Systems}
For high-dimensional or continuous state spaces, or for non-Markovian sequence models like Transformers, the analytic approach becomes intractable. In these cases, $\mathcal{H}$ must be estimated via Monte Carlo sampling. For each starting history $h_{start}$ in a training mini-batch, we generate $K$ independent future trajectories (rollouts) of length $\tau$ by autoregressively sampling actions from the policy. The estimator for the path-entropy functional is:
\begin{equation}
\label{eq:H_sampling_sup}
\hat{\mathcal{H}}_{\tau,\beta}(\theta) \approx \frac{1}{|\mathcal{B}|} \sum_{h_{start} \in \mathcal{B}} \left( \frac{1}{K\tau} \sum_{k=1}^{K} \sum_{j=0}^{\tau-1} \left[ -\ln \pi_{\theta,\beta}(a_{j}^{(k)}|h_{j}^{(k)}) \right]_{h_{start}} \right).
\end{equation}
To ensure that gradients can be backpropagated through the sampling process, especially for discrete action spaces, reparameterization techniques are required. A standard method is the Gumbel-Softmax trick \cite{Maddison2017}, which provides a continuous, differentiable approximation to the sampling procedure. Alternatively, the gradient of the entropic objective can be estimated using policy gradient methods like REINFORCE \cite{Williams1992}, though this often suffers from high variance.

\section{S3. Parametric Dependencies of the Intuition Phase}
\label{sec:si_dependencies}
The emergence of the fragile intuition phase is a critical phenomenon highly sensitive to the model, data, and learning protocol parameters. Below, we detail the key dependencies we investigated.

\subsection{A. Future Horizon $\tau$}
\begin{figure}[h!]
\centering
\includegraphics[width=0.90\columnwidth]{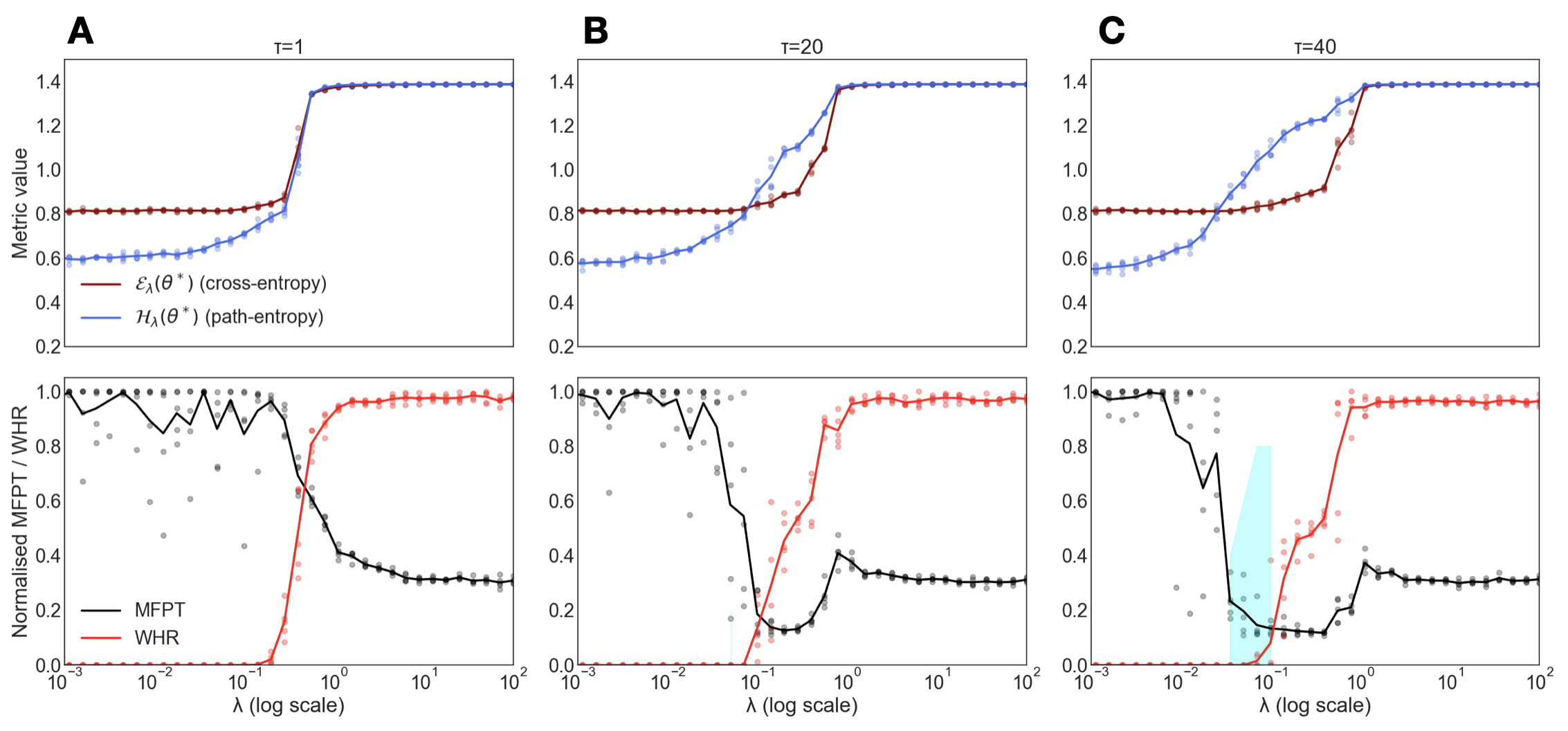}
\caption{\textbf{Dependence on Future Horizon $\tau$.} Phase diagram of the genotypic (top) and phenotypic (bottom) metrics as a function of $\lambda$ for different future horizons. The intuition window (sharp dip in MFPT and zero WHR, shaded blue) appears and stabilizes only for a long horizon ($\tau=40$). (A) A short horizon ($\tau=1$) yields only imitation and hallucination. (B) An intermediate horizon ($\tau=20$) can lead to a \emph{cheating} strategy, which is worse than the true intuitive solution (C).}
\label{fig:sup_horizon}
\end{figure}

The future horizon $\tau$ dictates the timescale of the model's ``imagination". Our experiments show that the intuition phase only emerges for a sufficiently long horizon (Fig. \ref{fig:sup_horizon}).

For a small $\tau$, the model is myopic; the long-term entropic gain from escaping the maze is not visible, so the model defaults to minimizing cross-entropy and remains in the imitation phase. As $\tau$ increases, the model can foresee the vast expansion of possible futures that awaits outside the maze, creating a strong entropic incentive to find an exit. For intermediate horizons, we often observe a \emph{cheating} phase—a local minimum in the free-energy landscape where the model learns to take a single illegal step through a wall. This strategy is a compromise: it incurs a small penalty for rule-breaking but gains a significant medium-term entropic advantage. Only for large $\tau$ does the incentive to find a legal path to maximal freedom dominate (i.e., virtue over vice).

\subsection{B. Model Capacity}
\begin{figure}[h!]
\centering
\includegraphics[width=0.90\columnwidth]{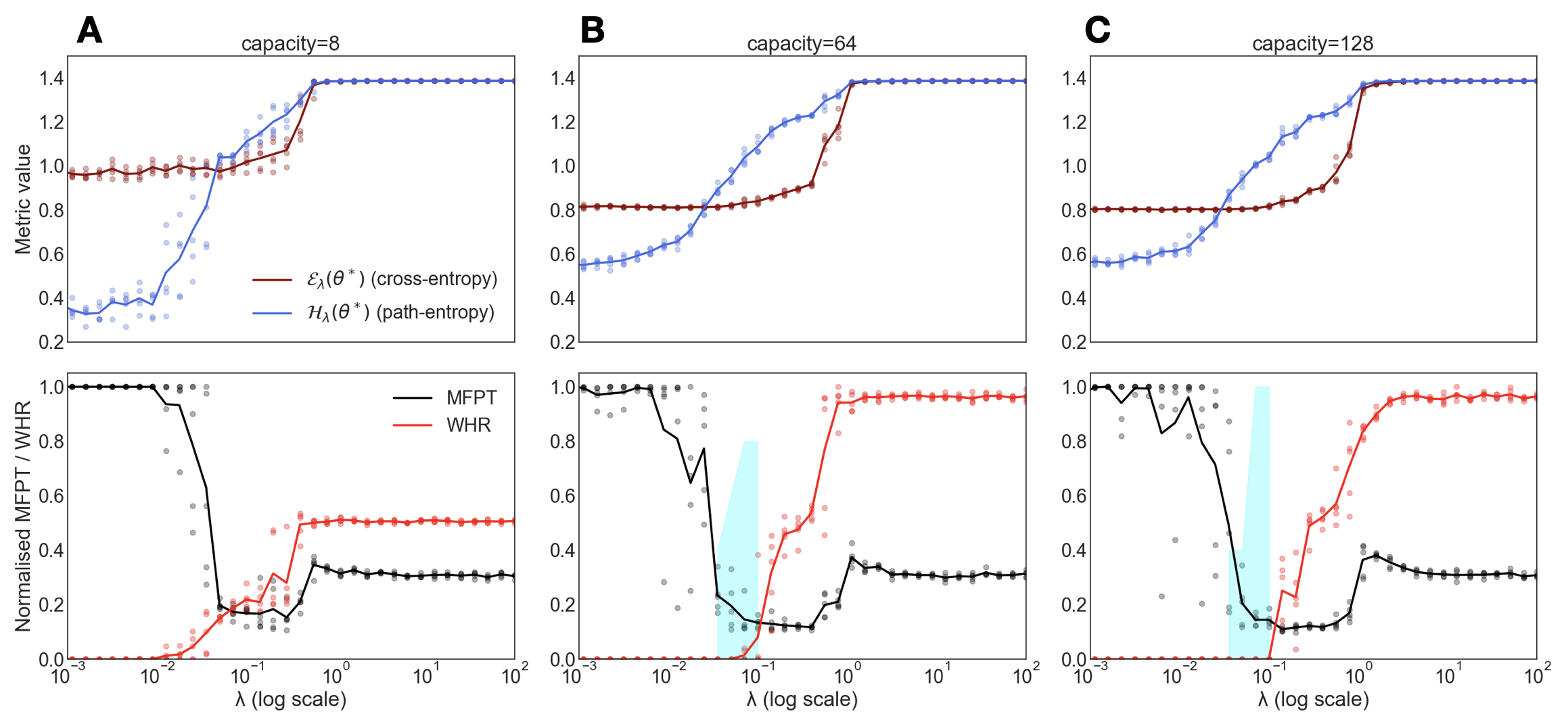}
\caption{\textbf{Dependence on Model Capacity.} Emergence of the intuition phase as a function of the number of neurons in the hidden layer. (A) A model with insufficient capacity (e.g., 8 neurons) cannot learn the required behavior. The intuition phase is robust for models with sufficient capacity (e.g., 64 (B) or 128 neurons (C)), which are powerful enough to discover the solution but not so powerful that they immediately overfit.}
\label{fig:sup_capacity}
\end{figure}

The capacity of the policy network, controlled by the number of neurons, is relevant (Fig. \ref{fig:sup_capacity}). A model with insufficient capacity has high bias and lacks the representational power to learn the complex, mixed strategy required to balance maze constraints with goal-directed exploration. It cannot simultaneously represent the world model and the entropic drive, so the intuition phase does not emerge. Conversely, a model with excessive capacity relative to the task complexity is prone to overfitting. It may perfectly memorize the noisy random walks from the training data or discover trivial, non-generalizable solutions (e.g., exploiting specific numerical artifacts) to maximize entropy. The intuition phase occupies a ``sweet spot" where model capacity is well-matched to the problem, enabling generalization from sparse data rather than mere memorization or unconstrained hallucination.

\subsection{C. Maze Complexity}
\begin{figure}[h!]
\centering
\includegraphics[width=\columnwidth]{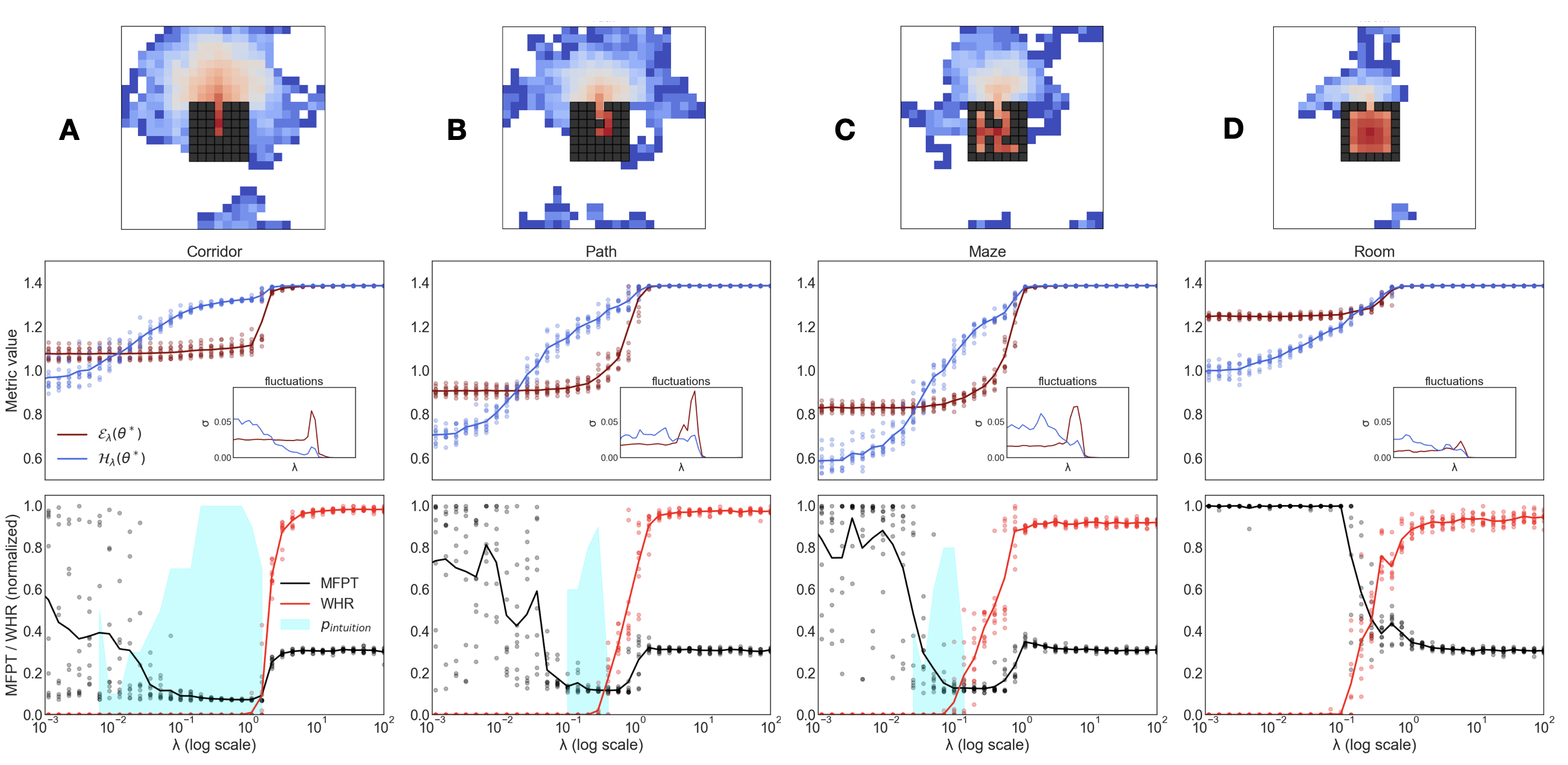}
\caption{\textbf{Dependence on Maze Complexity.} Position and width of the intuition window (measured by MFPT) for environments of varying difficulty. (\textbf{A,B}) For a simple corridor, the window is wide and appears at low $\lambda$. (\textbf{C}) For the more complex maze used in the main text, the window is narrower, reflecting the increased difficulty. (\textbf{D}) For even more complex problems, the intuition window can disappear, necessitating specific protocols to reach the desired phase.}
\label{fig:sup_complexity}
\end{figure}

We evaluated the framework on several environments of increasing complexity (Fig. \ref{fig:sup_complexity}). In simpler environments (e.g., a straight corridor), the escape task is trivial because the data trajectories are very close to the optimal solution. The intuition window is consequently wide and appears at lower values of $\lambda$. As maze complexity increases, finding the optimal path becomes a harder constraint-satisfaction problem. The cross-entropy term $\mathcal{E}$ more strongly penalizes deviations from valid paths. To overcome this, a stronger entropic pressure (a higher $\lambda$) is required to motivate the search for the distant exit. As a result, the intuition window narrows and shifts in the phase diagram, indicating that a more precise tuning of the energy-entropy balance is needed for more difficult problems. In some cases, the intuition window may disappear entirely, requiring protocols like the adiabatic sweep to be reached.

\section{S4. Detailed Theory and Further Predictions}
\label{sec:si_detailed_theory}

Here we expand on the theory from the main text, providing the explicit analytical forms for the free energy functional. We also clarify the calculation of the intuition likelihood ($p_{\text{intuition}}$) and discuss the existence of a more elusive \emph{inspiration phase} as a further prediction of the theory.

\subsection{A. The Effective Free Energy Functional}
For the Markovian maze environment with a small state-space, the terms of the free energy functional $\mathcal{F}_\lambda(m) = \mathcal{E}(m) - \lambda\mathcal{H}(m)$ can be computed analytically as a function of the rationality order parameter $m$. Note that this theory is effective: the $\beta$ of the analytical policy is distinct from the experimental one, since the former controls only three minimal analytical costs while the latter modulates the entire logit vector.

The effective cross-entropy, $\mathcal{E}(m)$, is the expectation of the negative log-likelihood over the state distribution of the given data, $\rho_{\mathcal{D}}(s)$. For a single state $s$, the cross-entropy is $E_s(m) = \log Z_m(s) + \beta \langle U_m(a) \rangle_{a \sim \text{uniform}}$, where the average is over legal moves from $s$. Summing over the data distribution gives
\begin{equation}
\label{eq:E_effective_sup}
\mathcal{E}(m) = \left\langle \log \left( \sum_{a \in \mathcal{A}} e^{-\beta U_m(a|s)} \right) + \beta \frac{\sum_{a' \in \mathcal{A}_{\text{legal}}(s)} U_m(a'|s)}{|\mathcal{A}_{\text{legal}}(s)|} \right\rangle_{s \sim \rho_{\mathcal{D}}},
\end{equation}
where $U_m(a|s)$ is the cost of action $a$ in state $s$ (which depends on whether the move is optimal, suboptimal, or a wall collision) and $\mathcal{A}_{\text{legal}}(s)$ is the set of valid moves from $s$.

The effective path-entropy, $\mathcal{H}(m)$, is the time-averaged Shannon entropy of the policy $p_{m,\beta}$ over trajectories of length $\tau$ starting from an initial state distribution $\rho_0$ (in our case, a single point at the maze start). It is calculated using the policy-dependent transition matrix $M_m$
\begin{equation}
\label{eq:H_effective_sup}
\mathcal{H}(m) = \frac{1}{\tau} \sum_{k=0}^{\tau-1} \left( \sum_{s \in \mathcal{V}} (\rho_k)_s \cdot h_m(s) \right),
\end{equation}
where $\vec{\rho}_k = (M_m)^k \vec{\rho}_0$ is the state occupancy vector at time $k$, and $h_m(s)$ is the local policy entropy at state $s$:
\begin{equation}
\label{eq:h_local_sup}
h_m(s) = -\sum_{a \in \mathcal{A}} p_{m,\beta}(a|s) \log p_{m,\beta}(a|s).
\end{equation}
These analytical expressions are used to generate the theoretical plots in the main text.

\subsection{B. Calculating the Intuition Likelihood ($p_{\text{intuition}}$)}
The intuition metric, visualized as the cyan region in the plots, quantifies the model's ability to spontaneously follow the optimal path. In experiments, this intuition likelihood is measured as the fraction of independent trials where the system displays the optimal solution at inference (minimal MFPT with zero WHR).

The same empirical criterion can be applied to the effective theory. More interestingly, the intuition likelihood can also be calculated analytically if the optimal route is known. We define it as the joint probability of generating the true shortest path to the exit, for a horizon of $q$ steps (where $q$ depends on the maze topology). Let the optimal path be the sequence of states $z^* = (s_0^*, s_1^*, \dots, s_q^*)$, where $s_0^*$ is the starting position, and let $a_t^*$ be the optimal action to transition from $s_t^*$ to $s_{t+1}^*$. The intuition likelihood for a given rationality level $m$ is:
\begin{equation}
\label{eq:p_intuition_sup}
p_{\text{intuition}}(m) = \prod_{t=0}^{q-1} p_{m,\beta}(a_t^* | s_t^*)
\end{equation}
Since the system can be multistable, the final value reported in the figure for a given $\lambda$ is the Boltzmann-weighted average of this likelihood over all coexisting free energy minima ($m^*, m^{**}, \dots$):
\begin{equation}
\label{eq:p_intuition_avg_sup}
\langle p_{\text{intuition}} \rangle_\lambda = \sum_{i} w_i(\lambda) \cdot p_{\text{intuition}}(m_i) \quad \text{where} \quad w_i(\lambda) = \frac{e^{-\hat{\beta}\mathcal{F}_\lambda(m_i)}}{\sum_j e^{-\hat{\beta}\mathcal{F}_\lambda(m_j)}},
\end{equation}
where $\hat{\beta}$ is an inverse temperature controlling the sampling of minima. At high $\hat{\beta}$ (low thermal noise), the system predominantly samples the global minimum, reproducing the steady-state results of the main experiments. In the numerical experiments, each run starts from a random weight initialization, and gradient descent acts as a local search that can fall into any of the attracting states. The likelihood metric is therefore zero in the imitation and hallucination phases (where the probability of following the optimal path is negligible) and peaks sharply in the narrow intuition window, provided the policy's inference $\beta$ is sufficiently high.

\subsection{C. From Intuition to Inspiration: Further Predictions of the Effective Theory}
The intuition phase represents a significant discovery, where the model finds a hidden, optimal solution that smoothly branches from the data-driven imitation phase. It is a better way, but not a radical departure. Intriguingly, the theory predicts the existence of a distinct, more profound cognitive phase: \emph{inspiration}. Inspiration is not a continuous improvement but an abrupt jump to a qualitatively different state of insight. This corresponds to the emergence of a new, globally optimal minimum in the free-energy landscape, where the rationality parameter is close to $m \approx 1$. A model in the inspiration phase does not merely approximate the optimal policy; it \emph{knows} the solution is correct. This internalized understanding would manifest through a key operational signature: the model could execute the optimal strategy robustly, even with a stochastic policy (low inference $\beta$), distinguishing it from the more tentative intuitive state.

\begin{figure}[h!]
\centering
\includegraphics[width=\columnwidth]{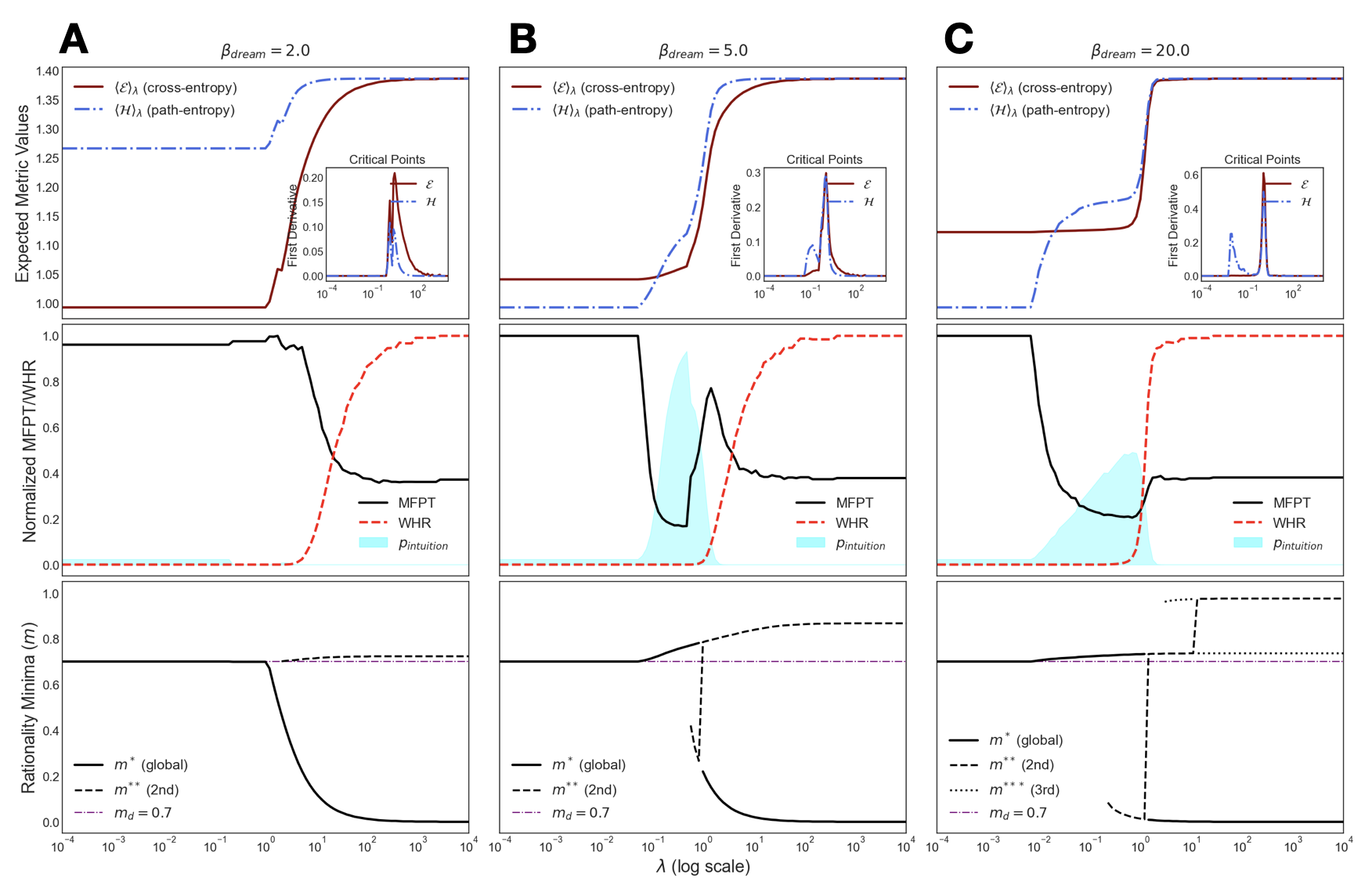}
\caption{\textbf{Dependence of theoretical predictions on the imagination temperature $\beta_{\text{dream}}$.} The theoretical phase diagram is shown for increasing values of $\beta_{\text{dream}} = \{2.0, 5.0, 20.0\}$. This parameter controls the policy stochasticity in the self-referential entropy calculation. \textbf{(A-C)} As $\beta_{\text{dream}}$ increases, the system's phase diagram (bottom row) changes. Higher values of this temperature can also reveal more complex phase structures, including the emergence of the inspiration phase, as discussed in the text. Insets in the first row (here and in the main text) measure the numerical first-derivatives of both the cross-entropy and path-entropy for low sampling temperature at equilibrium (thus for global attractors). The separation in the peaks of the discontinuities (B,C) signal the entropy- and energy- driven transitions that delimitate the intuition window.}
\label{fig:sup_betadream}
\end{figure}

The theory predicts that the imagination temperature $\beta_{\text{dream}}$—the policy stochasticity in the entropy term—is a key parameter for accessing these states (Fig.~\ref{fig:sup_betadream}). At low $\beta_{\text{dream}}$, the intuition phase ($m>m_D$) is unstable. It emerges in a stable window only for sufficiently large $\beta_{\text{dream}}$. At even higher values, this stable intuition branch can bifurcate into two locally stable solutions: the familiar \emph{intuition} phase and this hidden \emph{inspiration} phase ($m \approx 1$). Both can coexist while the hallucination phase ($m \ll m_D$) remains the global attractor. Observing this more exotic inspiration phase in practice would likely require careful tuning protocols, potentially starting from the intuition phase and employing non-equilibrium dynamics.

\end{document}